\documentclass[aps,pra,twocolumn,nofootinbib,longbibliography]{revtex4-2}

\usepackage{amsmath,amssymb,bm}
\usepackage{mathrsfs} 
\usepackage{graphicx}
\usepackage{xcolor}
\usepackage{hyperref}
\hypersetup{colorlinks=true,linkcolor=blue,citecolor=blue,urlcolor=blue}
\usepackage{times}
\newcommand{\Om}{\Omega}
\newcommand{\DelF}{\Delta_F}

\begin{document}
\author{Lu-Qi Yang}
\thanks{These authors contributed equally to this work.}
\author{Yu-Meng Ren}
\thanks{These authors contributed equally to this work.}
\author{Peng-Bo Li}
\email{lipengbo@mail.xjtu.edu.cn}
\affiliation{Ministry of Education Key Laboratory for Nonequilibrium Synthesis and Modulation of Condensed Matter, Shaanxi Province Key Laboratory of Quantum Information and Quantum Optoelectronic Devices, School of Physics, Xi'an Jiaotong University, Xi'an 710049, China}

\title{Nonreciprocal quantum rotation sensing via virtual-excitation enhancement in a spinning cavity}

\begin{abstract}
Quantum sensing with high precision and sensitivity plays an important role in quantum technologies and quantum information processing. Here, we propose a nonreciprocal quantum metrological scheme for estimating rotational angular velocity in a hybrid light-matter platform, where the setup consists of a spinning ring cavity coupled to a two-level system and an auxiliary bosonic mode. Through the Sagnac effect, the angular velocity is converted into a direction-dependent detuning, which modifies the effective light-matter dressing of the hybrid system. As a result, the angular velocity is encoded not only into the renormalized hybrid-mode spectrum, but also into the virtual excitations generated by ultrastrong coupling. These virtual excitations modify the polaritonic frequency response to rotation and enhance the quantum Fisher information (QFI) associated with angular velocity estimation, without requiring direct extraction of virtual excitations. Moreover, since the Sagnac-Fizeau shift  enters the virtual-transition energy denominators, the metrological response becomes intrinsically different for opposite driving directions, leading to a tunable nonreciprocal sensitivity contrast. In addition, we also discuss a readout scheme and show that bundle emission coincidence counting can serve as an auxiliary direction-dependent readout channel. Our results provide a route toward exploiting nonreciprocal light-matter dressing and virtual excitations as resources for quantum rotation sensing.

\end{abstract}

\maketitle

\section{\label{sec:intro}Introduction}

The central task of quantum precision measurement is to achieve high-precision estimation of unknown parameters with minimal resources in the inevitable presence of noise and decoherence~\cite{Giovannetti2006,Giovannetti2011,BraunsteinCaves1994,Helstrom1969,Helstrom1976,Abiuso2025,Degen2017,Pezze2018,Braun2018}. Conventional schemes typically rely on engineered techniques such as externally applied squeezing, entanglement, or measurement back-action evasion~\cite{Caves1981,Giovannetti2004,Pezze2009}, or exploit quantum criticality to enhance parameter sensitivity near phase 
transitions~\cite{Zanardi2008,Salvatori2014,Garbe2020,Rams2018,Salvia2023,PRL_CriticalMetrology_2024,Xue2026,Alushi2024,DiCandia2023,Ilias2022,Ding2022,Chu2021}.
In strongly coupled platforms, however, the intrinsic structure of the system is reshaped by interactions, generating virtual excitations and virtual squeezing as ground-state or steady-state resources~\cite{DeLiberato2017,Ciuti2005,GietkaHotterRitsch2023,
GietkaRitsch2023}. These virtual resources can enhance metrological sensitivity through the renormalized normal modes frequencies, without requiring direct extraction of the virtual excitations~\cite{PRL_USC_Metrology_2025}.
The ultrastrong coupling (USC) regime provides an important conceptual and experimental platform for realizing intrinsic virtual-excitation resources~\cite{Kockum2019,FornDiaz2019,Qin2024,Ciuti2005,Giannelli2024}. Related cavity-optomechanical and cold-atom cavity platforms offer experimentally relevant settings for engineering strong hybrid 
light-matter interactions~\cite{Aspelmeyer2014,Ritsch2013}.

On the other hand, nonreciprocal quantum processes provide key
resources for directional signal routing, quantum state transfer, and nonreciprocal control in diverse
platforms~\cite{Yao2025,Ren2025,Shen2023,Kannan2023,Huang2021,Xu2019},
as well as nonreciprocal synchronization~\cite{Lai2025}.
Spinning resonators introduce direction-dependent frequency
splitting via the Sagnac effect~\cite{Maayani2018,Lu2017},
and have been widely exploited to realize novel quantum effects and phenomena~\cite{Zhang2020,Tan2025,Xu2026,Lodahl2017}, including nonreciprocal photon or phonon blockade~\cite{Huang2018,Li2019,Yao2022}, optical nonreciprocity induced by quantum squeezing~\cite{Xia2022}, nonreciprocal quantum entanglement~\cite{Jiao2020,Chen2023PRB}, nonreciprocal
topological phonon transfer~\cite{Lai2024}, nonreciprocal
superradiant phase transitions~\cite{Fruchart2021,Zhu2024},
and the nonreciprocal Dicke model~\cite{Chiacchio2023}.
Recent work has further demonstrated that the frequency splitting induced by the Sagnac effect in a spinning resonator, combined with a
strongly driven two-level system, can realize switchable
nonreciprocal bundle emission of entangled photon-phonon and
photon-magnon pairs~\cite{PRL_NonreciprocalBundle_2024}.
Related works have also explored nonreciprocal phonon
lasing~\cite{Jiang2018phonon} and nonreciprocal photon
correlations induced by directional quantum
squeezing~\cite{Shen2023PRA}. 

However, the metrological potential of such a rotation-sensitive hybrid system has not been systematically explored within a quantum sensing framework. In this work, we formulate a hybrid platform consisting of a spinning ring cavity coupled to a two-level system and an auxiliary bosonic mode, as a nonreciprocal quantum sensor for estimating the rotational angular velocity. Building on the idea of exploiting the virtual excitations in ultrastrong coupling systems for quantum metrology, we investigate how this metrological resource can be combined with rotation-induced nonreciprocity. These virtual excitations can modify the polaritonic response to rotation and thereby enhance the quantum Fisher information, without requiring externally prepared squeezed probes or direct extraction of virtual excitations. The advantage of our scheme is that this virtual-excitation metrological response becomes direction dependent, giving rise to different sensitivities for the opposite driving configurations. We emphasize that rotation-induced nonreciprocity plays an important role in this mechanism. In a spinning ring cavity, the Sagnac-Fizeau frequency shift $\DelF(\Om)$ produces a direction-dependent frequency shift of the cavity resonance, as used in the conventional rotating sensing scheme. However, for the hybrid system considered here, it also modifies the effective detuning that governs the light-matter dressing. It affects not only the bare cavity resonance frequency but also the virtual-transition processes responsible for the renormalization of the effective cavity frequency and light-matter coupling. Since the two counter-propagating driving configurations experience Sagnac-Fizeau shifts with opposite signs, they follow distinct trajectories in the effective-parameter space and exhibit different metrological responses. Therefore, the angular velocity $\Om$ is encoded not only in the shifted cavity resonance, but also in the renormalized hybrid-mode spectrum and the virtual-excitation structure of the ultrastrongly coupled hybrid system. The asymmetric response between opposite directions enables optimization of the operating point toward the more sensitive driving direction and provides a differential rotation-direction readout. We quantify the rotation sensitivity using the quantum Fisher information and discuss experimentally accessible readout strategies, with the lower-polariton spectral response serving as the primary readout and bundle-emission coincidence counting serving as an auxiliary direction-selective readout channel. This work shows the joint use of virtual-excitation resources and nonreciprocal directional control as a general strategy for enhancing direction-sensitive quantum metrology in hybrid quantum systems.

\section{\label{sec:model}Theoretical Model}

We consider a spinning ring optical microcavity supporting two 
counter-propagating optical modes, the clockwise (CW) and 
counter-clockwise (CCW) modes. Rotation induces opposite 
Sagnac-Fizeau shifts for the two modes. Throughout this paper 
$\DelF(\Om)>0$ denotes the positive shift magnitude~\cite{Post1967,Malykin2000,Maayani2018}
\begin{equation}
\DelF(\Om)=\frac{Rn_R\omega_a\Om}{c}
\left[1-\frac{1}{n_R^2}-\frac{\lambda}{n_R}\frac{dn_R}{d\lambda}\right],
\label{eq:fizeau_mag}
\end{equation}
where $R$ is the cavity radius, $n_R$ is the refractive index, 
$\lambda$ is the vacuum wavelength, and $\Om>0$ is the rotation 
angular velocity. The drive direction is encoded by $s=+1$ (LD) 
and $s=-1$ (RD), so that the direction-dependent cavity resonance is
\begin{equation}
\omega_{a}^{(s)}=\omega_a+s\DelF(\Om),
\label{eq:fizeau_shift}
\end{equation}
which differs in sign convention from Ref.~\cite{PRL_NonreciprocalBundle_2024} 
but is physically equivalent. The estimation of $\Om$ is thus 
equivalent to estimating the direction-dependent detuning 
$\Delta_s(\Om)=\Delta_{ad}+s\DelF(\Om)$, which is the central 
resource for rotation metrology in the present scheme.

\begin{figure}[t]
  \centering
  \includegraphics[width=\linewidth]{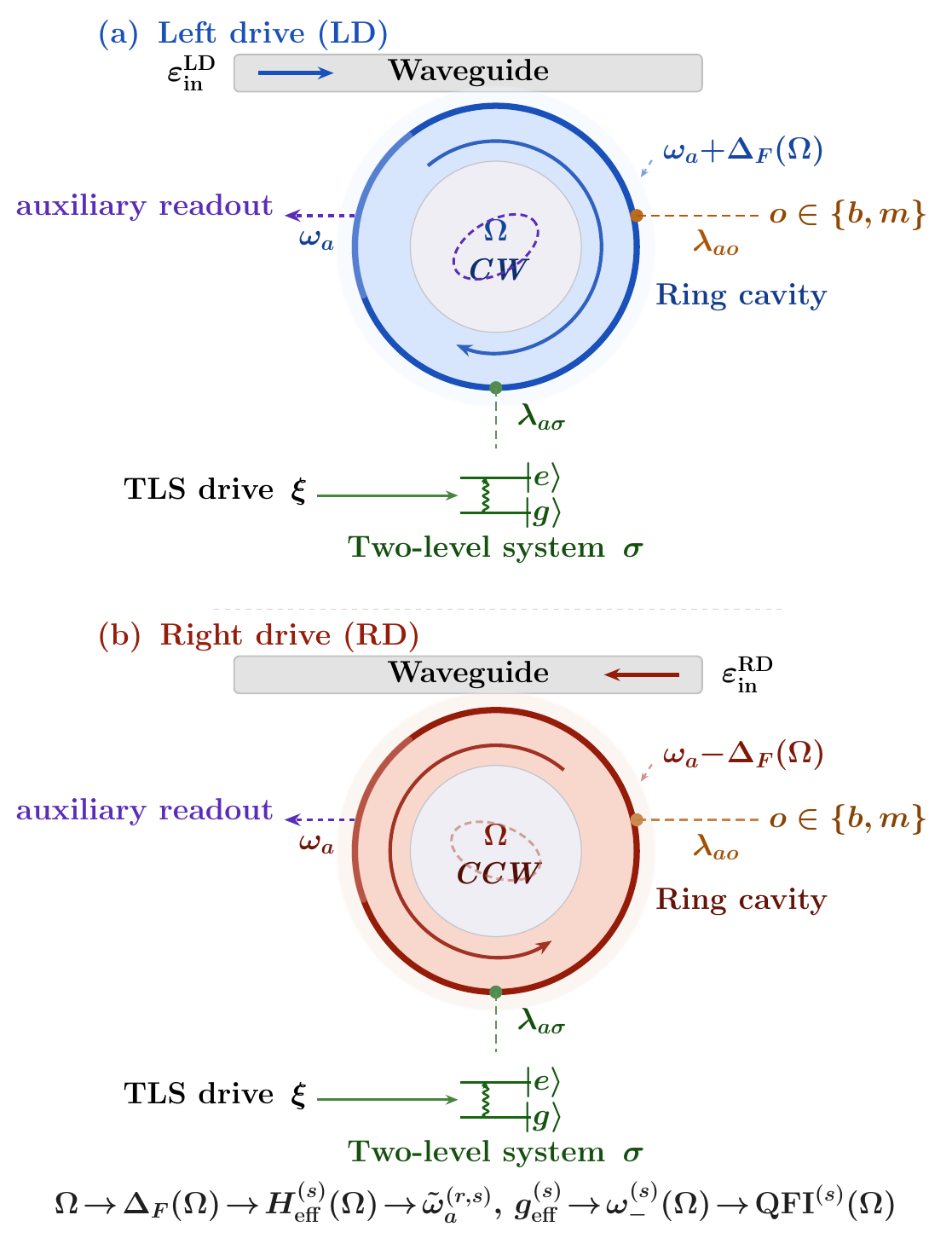}
  \caption{Schematic of the hybrid quantum sensing system.
  A spinning ring microresonator is coupled to a driven two-level system (TLS, $\sigma$) via $\lambda_{a\sigma}$ and to a bosonic mode $o\in\{b,m\}$ via the linearized coupling
  $\lambda_{ao}$.
  \textbf{(a)~Left drive (LD):} The CW optical mode is excited,
  shifting the cavity resonance to $\omega_a+\Delta_F(\Omega)$.
  \textbf{(b)~Right drive (RD):} The CCW optical mode is excited,
  shifting the cavity resonance to $\omega_a-\Delta_F(\Omega)$.
  The two drive directions therefore realize different effective
  detuning configurations and follow distinct parameter trajectories
  in the SW-Bogoliubov effective Hamiltonian.
  Depending on the selected resonance condition, bundle emission
  signals may provide an auxiliary direction-selective readout
  channel, while the primary metrological response is obtained from
  the lower-polariton spectral sensitivity.
  The bottom arrow summarizes the sensing chain:
  $\Omega\!\to\!\Delta_F(\Omega)\!\to\!
  H_\mathrm{eff}^{(s)}(\Omega)
  \!\to\!\tilde{\omega}_a^{(r,s)},\,g_\mathrm{eff}^{(s)}
  \!\to\!\omega_-^{(s)}(\Omega)
  \!\to\!\mathrm{QFI}^{(s)}(\Omega)$.}
  \label{fig:model}
\end{figure}

As schematically shown in Fig.~\ref{fig:model}, the system contains a coherently and strongly driven two-level system (TLS) with lowering operator $\sigma=|g\rangle\langle e|$ coupled to the optical mode of the spinning cavity with annihilation operator $a$. We denote a bosonic mode (either a mechanical (phonon) mode or a magnon mode) by $o$ with the frequency $\omega_o$ and a linearized coupling strength $\lambda_{ao}$ formed by the cavity field. After transforming to the rotating frame of the drive frequency $\omega_d$ and applying the standard linearization for the coupling between the cavity and the bosonic mode $o$, the effective system Hamiltonian reads ($\hbar=1$)~\cite{PRL_NonreciprocalBundle_2024,Aspelmeyer2014,ScullyZubairy}
\begin{align}
H
=&\;\Delta_s(\Omega)a^\dagger a+\omega_o o^\dagger o
+\Delta_{\sigma d}\sigma^\dagger\sigma\nonumber\\
&+\lambda_{ao}(a^\dagger+a)(o^\dagger+o)+\lambda_{a\sigma}(a\sigma^\dagger+a^\dagger\sigma)+\xi(\sigma^\dagger+\sigma),
\label{eq:full_H}
\end{align}
where $\Delta_s(\Omega)=\Delta_{ad}+s\DelF(\Omega)$ with $\Delta_{ad}=\omega_a-\omega_d$ and
$\Delta_{\sigma d}=\omega_\sigma-\omega_d$ are the optical mode and TLS detunings from the drive, respectively. Here, $\lambda_{a\sigma}$ is the cavity-TLS coupling, and $\xi$ is the coherent drive amplitude applied on the TLS.

With strong drive $\xi$ and detuning $\Delta_{\sigma d}$, the driven TLS Hamiltonian can be written as $H_{\rm TLS}=\Delta_{\sigma d}\sigma^\dagger\sigma+\xi(\sigma^\dagger+\sigma)$. The dressed eigenenergies are~\cite{ScullyZubairy}:
\begin{equation}
E_\pm=\frac{\Delta_{\sigma d}}{2}\pm\frac{1}{2}\sqrt{\Delta_{\sigma d}^2+4\xi^2}.
\label{eq:dressed_energy}
\end{equation}
And we define the energy splitting of the dressed state:
\begin{equation}
    \omega_q=E_+-E_-=\sqrt{\Delta_{\sigma d}^2+4\xi^2}.
\end{equation}

The Sagnac-shifted virtual transition denominators are defined as
\begin{align}
\Delta_\pm(\Om)
&=\Delta_s(\Om)\pm\omega_q.
\label{eq:delta_pm_main}
\end{align}
The dispersive regime required for the virtual-process-based metrological enhancement is
\begin{equation}
\left|\Delta_\pm(\Om)\right|
\gg \lambda_{a\sigma},
\label{eq:dispersive_condition}
\end{equation}
We now turn to obtain the effective two-mode Hamiltonian for the bosonic mode. Firstly, we eliminate the TLS by applying the Schrieffer-Wolff transformation~\cite{Bravyi2011}, followed by a single-mode Bogoliubov transformation on the cavity mode [see Appendix~\ref{app:geff_sw}].
Introducing the dressed states: 
\begin{align}
    &|+\rangle=\cos\theta |g\rangle+\sin\theta |e\rangle,\nonumber\\
    &|-\rangle=\sin\theta |g\rangle-\cos\theta |e\rangle,
\end{align}
with $\tan(2\theta)=2\xi/\Delta_{\sigma d}$, the cavity-TLS interaction is:
\begin{equation}
\begin{aligned}
H_{a\sigma}
=&\;\lambda_{a\sigma}\Big[
g_R(a\tau_+ +a^\dagger\tau_-)
-g_{CR}(a\tau_-+a^\dagger\tau_+) \\
&\qquad\qquad
+g_z(a+a^\dagger)\tau_z
\Big].
\end{aligned}
\end{equation}
where $g_R=\sin^2\theta$, $g_{CR}=\cos^2\theta$ and $g_z=\sin\theta\cos\theta$. And we define $\tau_+=|+\rangle\langle -|$, $\tau_-=|-\rangle\langle +|$ and $\tau_z=|+\rangle\langle +| - |-\rangle\langle -|$.
Here, the longitudinal term proportional to $g_z$ results to a static displacement after projection onto a fixed dressed branch and is removed from the Hamiltonian. The flip terms proportional to $g_{R}$ and $g_{CR}$ generate the second-order virtual processes, renormalizing the cavity mode.
After projecting onto the lower dressed branch $|-\rangle$ and applying the Schrieffer-Wolff transformation, we obtain the effective single cavity mode Hamiltonian:
\begin{equation}
    H_{a,{\rm SW}}^{(s)}(\Omega)=\tilde\omega_a^{(s)}(\Omega)a^\dagger a+\frac{\chi^{(s)}(\Omega)}{2}(a^2+a^{\dagger 2}),
    \label{eq:H_a,SW}
\end{equation}
where $\tilde\omega_a^{(s)}$ is Lamb-shifted cavity detuning induced by SW transformation and $\chi^{(s)}$ is the SW-induced single-mode squeezing coefficient. Using the denominator convention in Eq.~\eqref{eq:delta_pm_main}, we can obtain:
\begin{align}
\tilde\omega_a^{(s)}(\Om)
&=\Delta_s(\Om)+\lambda_{a\sigma}^2\!\left[
\frac{g_R^2}{\Delta_-(\Om)}-\frac{g_{CR}^2}{\Delta_+(\Om)}
\right],
\label{eq:omega_a_tilde_main}\\
\chi^{(s)}(\Om)
&=-\lambda_{a\sigma}^2\,g_Rg_{CR}\!\left[
\frac{1}{\Delta_-(\Om)}-\frac{1}{\Delta_+(\Om)}
\right].
\label{eq:chi_def_main}
\end{align}

The Hamiltonian in Eq.~\eqref{eq:H_a,SW} can be diagonalized by the single-mode Bogoliubov transformation $a=\cosh{r^{(s)}}\tilde{a} +\sinh{r^{(s)}}\tilde{a}^\dagger$, where the squeezing parameter is determined by $\tanh{[2r^{(s)}(\Omega)]}=-\chi^{(s)}(\Omega)/{\tilde\omega_a^{(s)}(\Omega)}$ and the cavity mode is mapped to $\tilde a$. Thus, the effective two-mode Hamiltonian renormalized by the Bogoliubov transformation becomes
\begin{equation}
H_{\rm eff}^{(s)}(\Om)
=\tilde\omega_a^{(r,s)}(\Om)\,\tilde{a}^\dagger\tilde{a}
+\omega_o\,o^\dagger o
+g_{\rm eff}^{(s)}(\Om)\,(\tilde{a}+\tilde{a}^\dagger)(o+o^\dagger),
\label{eq:Heff}
\end{equation}
where
$\tilde\omega_a^{(r,s)}(\Om)
=\tilde\omega_a^{(s)}(\Om)/\cosh[2r^{(s)}(\Om)]$
is the doubly renormalized cavity frequency, with the SW-induced $\tilde\omega_a^{(s)}(\Om)$ given by Eq.~\eqref{eq:omega_a_tilde_main} and $r^{(s)}(\Om)$ defined by the Bogoliubov transformation. At the same time, the effective coupling between the cavity and bosonic mode $o$ is renormalized with:
\begin{equation}
g_{\rm eff}^{(s)}(\Om)=\lambda_{ao}\,e^{r^{(s)}(\Om)},
\label{eq:geff}
\end{equation}
which is enhanced by the Bogoliubov squeezing parameter $r^{(s)}(\Om)$. In the weak-coupling limit $\lambda_{a\sigma}\to0$, one obtains
$g_{\rm eff}^{(s)}\to\lambda_{ao}$, recovering the bare coupling. The derivations of the above transformations are detailed in Appendix~\ref{app:geff_sw}.

\begin{figure}[t]
  \centering
  \includegraphics[width=\linewidth]{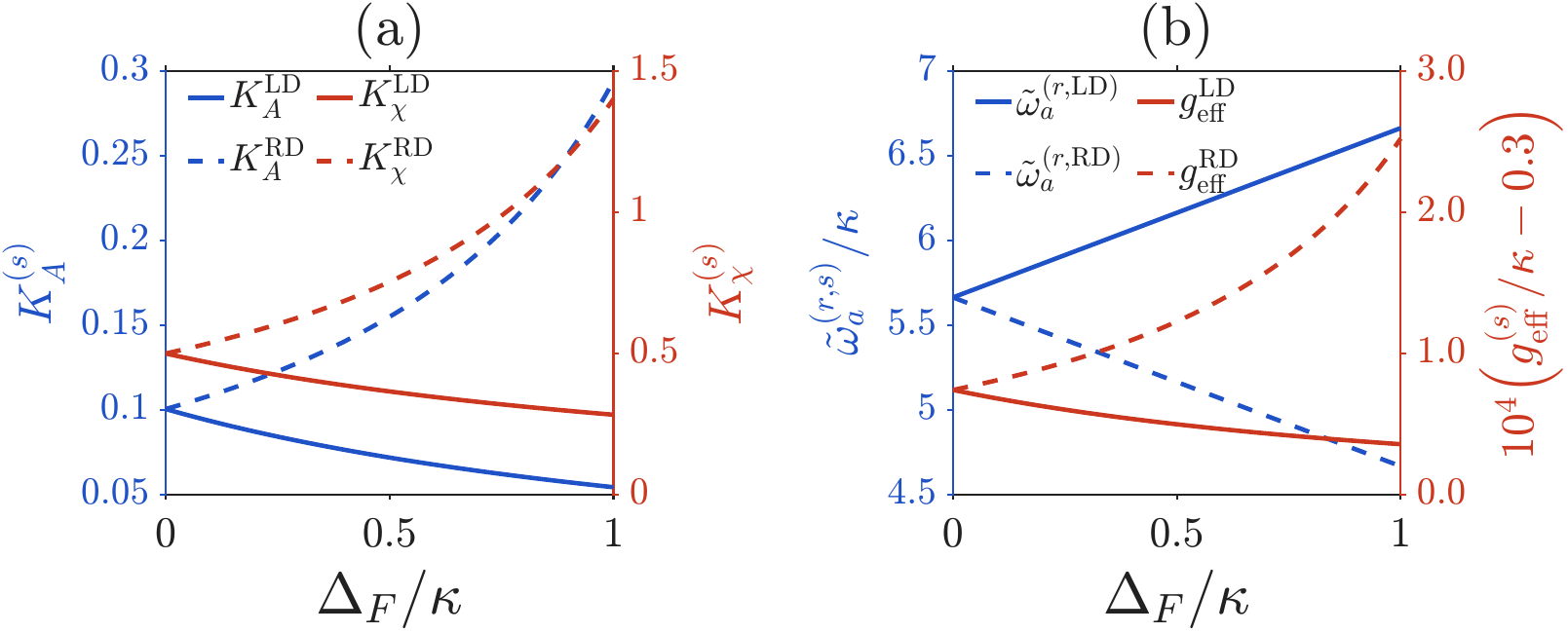}
  \caption{
SW-Bogoliubov effective parameter response to the Sagnac-Fizeau
shift, with $\Delta_{ad}/\kappa=5.66$.
(a)~The Schrieffer-Wolff kernels
$K_A^{(s)}=g_R^2/\Delta_- - g_{CR}^2/\Delta_+$
and
$K_\chi^{(s)}=1/\Delta_- - 1/\Delta_+$
as functions of $\Delta_F$ for LD (solid) and RD (dashed).
The plotted kernel values are normalized by $\kappa$, i.e.,
the vertical axes show the dimensionless values of the kernels
in units of $\kappa^{-1}$.
The denominators $\Delta_\pm$ are defined in
Eq.~\eqref{eq:delta_pm_main}.
(b)~The corresponding Bogoliubov-dressed cavity frequency
$\tilde{\omega}_a^{(r,s)}$ and effective coupling
$g_{\rm eff}^{(s)}$. For visual clarity, the right vertical axis
shows $10^4[g_{\rm eff}^{(s)}/\kappa-0.3]$.
}
  \label{fig:sw_bogoliubov_response}
\end{figure}

It is worth emphasizing that the physical difference between the
LD and RD responses is not merely a formal sign flip of
$\DelF(\Om)$. Once $\DelF(\Om)$ enters the Schrieffer-Wolff
denominators $\Delta_\pm(\Om)=\Delta_s(\Om)\pm\omega_q$, it changes
the two branch-dependent second-order kernels
$K_A^{(s)}=g_R^2/\Delta_--g_{CR}^2/\Delta_+$
and
$K_\chi^{(s)}=1/\Delta_--1/\Delta_+$.
These kernels determine the SW frequency corrections and the
TLS-mediated virtual squeezing terms, respectively. Subsequently, these kernels are converted by the Bogoliubov transformation into both the
Bogoliubov-dressed cavity frequency $\tilde{\omega}_a^{(r,s)}(\Om)$ and the effective coupling $g_{\rm eff}^{(s)}(\Om)$. 
As illustrated in Fig.~\ref{fig:sw_bogoliubov_response}, we show how these intermediate conversions from the direction-dependent SW kernels enter to the effective parameters. The final spectral
sensitivity is then obtained only after these effective parameters
entering the lower-polariton normal-mode response, as analyzed in
Sec.~\ref{sec:qfi}.

The interaction term in Eq.~\eqref{eq:Heff} originates from the
position--position coupling
$\lambda_{ao}(a+a^\dagger)(o+o^\dagger)$ in Eq.~\eqref{eq:full_H}.
It contains the counter-rotating components
$\tilde a^\dagger o^\dagger$ and $\tilde a o$, which allow the
coupled ground state to acquire virtual-excitation occupation without
an externally applied squeezing drive~\cite{Ciuti2005,Emary2003}.
This provides intrinsic quantum resources for the subsequent
sensitivity enhancement.

\section{\label{sec:qfi}Quantum Fisher Information and Nonreciprocal Sensitivity Enhancement}

Taking the rotation angular velocity $\Om$ as the parameter to be
estimated, the precision of any unbiased estimator satisfies the
quantum Cram\'{e}r--Rao bound~\cite{BraunsteinCaves1994,Helstrom1976,Rao1992}
\begin{equation}
\delta\Om\geq\frac{1}{\sqrt{\nu F_Q^{\rm exact}(\Om)}},
\label{eq:QCRB}
\end{equation}
where $\nu$ is the number of independent experimental runs.
Since $H_{\rm eff}^{(s)}(\Om)$ is quadratic and the initial state
is Gaussian, $F_Q^{\rm exact}(\Om)$ is determined by the covariance
matrix and displacement vector of the two-mode Gaussian state
[Appendix~\ref{app:gaussian_qfi}, Eq.~\eqref{eq:QFI_Gaussian}].

The proxy QFI is constructed by projecting the full two-mode
response onto the lower-polariton branch. When the two polariton
branches are spectrally well separated and the coherent probe is
tuned near $\omega_-^{(s)}(\Om)$, the first-moment response of
the two-mode Gaussian state is dominated by the lower-polariton
mode $\hat{e}$ [see Appendix~\ref{app:gaussian_qfi}], and
the proxy QFI takes the form
\begin{equation}
F_{Q,\mathrm{proxy}}^{(s)}(\Om)
=
4t^2|\alpha|^2
\left[
\partial_\Om\omega_-^{(s)}(\Om)
\right]^2.
\label{eq:FQ_proxy}
\end{equation}
This expression approximates $F_Q^{\rm exact}(\Om)$ when
$|\alpha|^2\gg1$. In the reduced single-polariton limit
$H_-^{(s)}=\omega_-^{(s)}\hat{e}^\dagger\hat{e}$,
Eq.~\eqref{eq:FQ_proxy} becomes exact
[see Appendix~\ref{app:gaussian_qfi}].
For general parameter ranges, $\partial_\Omega\omega_-^{(s)}$ is
evaluated using the closed-form lower-polariton derivative derived in Appendix~\ref{app:diag_and_derivative}, Eq.~\eqref{eq:dw_result_app},
which includes both
$\partial_\Omega\tilde{\omega}_a^{(r,s)}$ and
$\partial_\Omega g_{\rm eff}^{(s)}$. In the near-critical regime
$q_s\to1^-$, an analytic approximation can be obtained as follows.

\subsection{\label{sec:qfi-A}Polaritonic softening threshold}
The effective Hamiltonian $H_{\rm eff}^{(s)}$ in Eq.~\eqref{eq:Heff} contains the
doubly renormalized cavity frequency $\tilde{\omega}_a^{(r,s)}$
and the effective coupling
$g_{\rm eff}^{(s)}=\lambda_{ao}e^{r^{(s)}}$, both derived in
Appendix~\ref{app:geff_sw} under the dispersive
condition in Eq.~\eqref{eq:dispersive_condition}. Here $r^{(s)}(\Om)$
is the TLS-mediated single-mode squeezing parameter.
The near-threshold metrological enhancement is governed by the softening
$\omega_-^{(s)}\to0^+$ of the physical lower-polariton branch.
From the Hopfield/Bogoliubov diagonalization of
$H_{\rm eff}^{(s)}$ [see Appendix~\ref{app:diag_and_derivative}],
the condition $\omega_-^{(s)}(\Om)=0$ gives the
direction-dependent critical coupling
\begin{equation}
g_{{\rm eff},c}^{(s)}(\Om)
=
\frac{1}{2}
\sqrt{\tilde{\omega}_a^{(r,s)}(\Om)\,\omega_o}.
\label{eq:geff_c_qfi}
\end{equation}
We therefore define the dimensionless critical-distance parameter
\begin{equation}
q_s(\Om)
\equiv
\frac{g_{\rm eff}^{(s)}(\Om)}{g_{{\rm eff},c}^{(s)}(\Om)}
=
\frac{2g_{\rm eff}^{(s)}(\Om)}
{\sqrt{\tilde{\omega}_a^{(r,s)}(\Om)\,\omega_o}}.
\label{eq:q_s_qfi}
\end{equation}
The normal-phase region corresponds to $q_s<1$, while
$0<1-q_s\ll1$ defines the near-critical sensing regime.

Substituting the explicit expressions for
$\tilde{\omega}_a^{(r,s)}$ and $g_{\rm eff}^{(s)}$ from
Appendix~\ref{app:geff_sw} into Eq.~\eqref{eq:q_s_qfi},
the factors involving
$[\tilde{\omega}_a^{(s)}-\chi^{(s)}]$ cancel, yielding
\begin{equation}
q_s(\Om)
=
\frac{2\lambda_{ao}}
{\sqrt{\omega_o\!\left[
\tilde{\omega}_a^{(s)}(\Om)+\chi^{(s)}(\Om)
\right]}}.
\label{eq:q_s_SW_qfi}
\end{equation}
The direction-dependent critical linearized coupling is therefore
\begin{equation}
\lambda_{ao,c}^{(s)}(\Om)
=
\frac{1}{2}
\sqrt{\omega_o\!\left[
\tilde{\omega}_a^{(s)}(\Om)+\chi^{(s)}(\Om)
\right]},
\label{eq:lambda_ao_c_qfi}
\end{equation}
with $q_s(\Om)=\lambda_{ao}/\lambda_{ao,c}^{(s)}(\Om)$.
Equation~\eqref{eq:lambda_ao_c_qfi} is analogous to the
critical-coupling condition in ultrastrong coupling quantum
metrology~\cite{PRL_USC_Metrology_2025,Garbe2020}.
In the present system, however, the critical threshold is jointly
renormalized by the TLS-mediated virtual quadratic process and the
direction-dependent Sagnac-Fizeau shift. Since $\DelF(\Om)$ enters the
SW denominators $\Delta_\pm(\Om)$, both
$\tilde{\omega}_a^{(r,s)}(\Om)$ and $g_{\rm eff}^{(s)}(\Om)$
become direction dependent, giving
\begin{equation}
\lambda_{ao,c}^{\rm LD}(\Om)\neq\lambda_{ao,c}^{\rm RD}(\Om),
\qquad
q_{\rm LD}(\Om)\neq q_{\rm RD}(\Om).
\label{eq:nonreciprocal_threshold}
\end{equation}
Thus, the LD and RD branches generally follow different
trajectories relative to the lower-polariton softening threshold,
producing a nonreciprocal sensitivity contrast.

\subsection{\label{sec:qfi-B}Near-critical analytic approximation}
From the polariton diagonalization [Appendix~\ref{app:diag_and_derivative}],
expanding $\omega_-^{(s)}(\Om)$ near $q_s\to1^-$ gives
\begin{equation}
\omega_-^{(s)}(\Om)
\simeq
\mathcal{W}^{(s)}(\Om)\sqrt{1-q_s(\Om)},
\label{eq:omega_minus_asymptotic_qfi}
\end{equation}
where
\begin{equation}
\mathcal{W}^{(s)}(\Om)
\equiv
\frac{\sqrt{2}\,\tilde{\omega}_a^{(r,s)}(\Om)\,\omega_o}
{\sqrt{[\tilde{\omega}_a^{(r,s)}(\Om)]^2+\omega_o^2}}.
\label{eq:W_eff_qfi}
\end{equation}
To expose the critical scaling, we introduce the lower-polariton
soft-mode parameter
\begin{equation}
r_-^{(s)}(\Om)
=
\frac{1}{4}\ln\!\left[1-q_s(\Om)\right]\leq0,
\label{eq:r_minus_qfi}
\end{equation}
so that Eq.~\eqref{eq:omega_minus_asymptotic_qfi} becomes
$\omega_-^{(s)}\simeq\mathcal{W}^{(s)}e^{2r_-^{(s)}}$.
In the symmetric near-resonant limit
$\tilde{\omega}_a^{(r,s)}\simeq\omega_o$,
$r_-^{(s)}$ reduces to the lower-polariton virtual squeezing
parameter used in ultrastrong coupling metrology~\cite{PRL_USC_Metrology_2025}.
Here $r^{(s)}$ denotes the TLS-mediated cavity mode squeezing
parameter, whereas $r_-^{(s)}$ characterizes the softening of
the physical lower-polariton branch.

Differentiating Eq.~\eqref{eq:omega_minus_asymptotic_qfi} and
retaining the dominant term as $q_s\to1^-$ gives
\begin{equation}
\partial_\Om\omega_-^{(s)}
\simeq
-\frac{\mathcal{W}^{(s)}\,\partial_\Om q_s}{2}
e^{-2r_-^{(s)}},
\label{eq:domega_exp_qfi}
\end{equation}
provided $\partial_\Om q_s(\Om)\neq0$.
Substituting into Eq.~\eqref{eq:FQ_proxy} and defining
$R_-^{(s)}\equiv-r_-^{(s)}>0$ yields
\begin{equation}
F_{Q,\mathrm{proxy}}^{(s)}(\Om)
\simeq
t^2|\alpha|^2
\left[\mathcal{W}^{(s)}\right]^2
\left[\partial_\Om q_s\right]^2
e^{4R_-^{(s)}(\Om)}.
\label{eq:FQ_exp_qfi}
\end{equation}
The exponential factor $e^{4R_-^{(s)}}$ originates from
lower-polariton softening, while the prefactor
$[\mathcal{W}^{(s)}]^2[\partial_\Om q_s]^2$ carries the
direction-dependent parameter response.

The numerical maps below retain parameter points satisfying the
dispersive condition in Eq.~\eqref{eq:dispersive_condition}, the
Bogoliubov stability condition
$\tilde{\omega}_a^{(s)}(\Om)>|\chi^{(s)}(\Om)|$,
and the normal-phase condition $q_s(\Om)<1$.
The additional condition $0<1-q_s\ll1$ specifies the subset in
which the asymptotic expression in  Eq.~\eqref{eq:FQ_exp_qfi} is
quantitatively applicable.
A direct numerical benchmark of this leading asymptotic expression
against the exact lower-polariton derivative used in the proxy QFI
calculation is given in Appendix~\ref{app:critical_validation},
confirming its convergence as $q_s\to1^-$.

\subsection{\label{sec:qfi-C}Direction-selective numerical evaluation}
\begin{figure}[t]
  \centering
  \includegraphics[width=\linewidth]{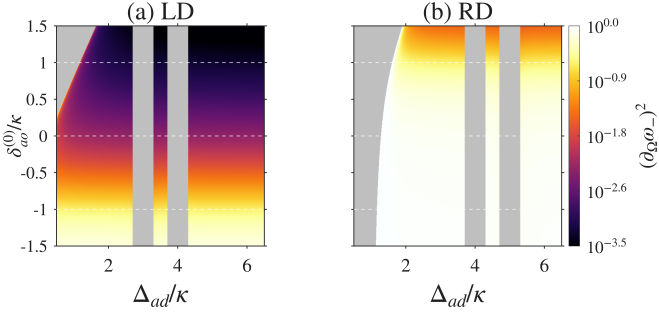}
  \caption{Direction-dependent lower-polariton sensitivity landscape.
  The color scale shows the normalized proxy QFI
  $F_{Q,\mathrm{proxy}}^{(s)}/(4t^2|\alpha|^2)
  =[\partial_{\Omega}\omega_-^{(s)}]^2$
  as a function of $\Delta_{ad}/\kappa$ and
  $\delta_{ao}^{(0)}/\kappa$.
  Panels (a) and (b) correspond to the LD and RD branches,
  respectively.
  Gray regions are excluded by the validity conditions of the
  effective model, and the horizontal dashed lines denote the cuts
  used in Fig.~\ref{fig:QFI_nonreciprocal}.
  For the parameters considered here, the RD branch exhibits a
  broader high-sensitivity region than the LD branch.}
  \label{fig:QFI_heatmap}
\end{figure}
All effective quantities are evaluated from the SW-induced cavity renormalization
and the subsequent Bogoliubov transformation in
Appendix~\ref{app:geff_sw}. The full proxy QFI maps are obtained
from Eq.~\eqref{eq:FQ_proxy} using the exact lower-polariton
derivative in Eq.~\eqref{eq:dw_result_app}, which contains both the
renormalized-frequency contribution
$\partial_\Om\tilde{\omega}_a^{(r,s)}$ and the effective-coupling
contribution $\partial_\Om g_{\rm eff}^{(s)}$. Here we use
$\delta_{ao}^{(0)}\equiv A_b^{(0)}-\omega_o$ to denote the
zero-rotation detuning between the Bogoliubov-renormalized cavity
mode and the bosonic mode $o$, with $A_b^{(0)}$ evaluated at
$\DelF=0$.

Figure~\ref{fig:QFI_heatmap} displays the proxy QFI landscapes for the two drive directions on the same physical parameter plane.
The opposite Sagnac-Fizeau shifts lead to distinct directional distributions. From the comparison between Fig.~\ref{fig:QFI_heatmap}(a) and (b), the RD branch accesses a broader high-sensitivity region for the representative parameters used here.

\begin{figure}[t]
  \centering
  \includegraphics[width=\linewidth]{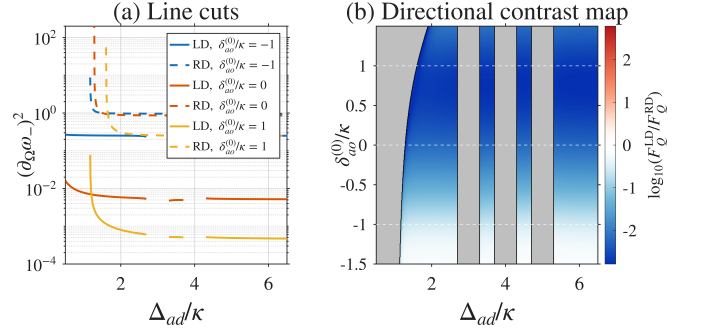}
  \caption{Directional comparison of the lower-polariton proxy QFI.
  (a)~Line cuts of the normalized proxy QFI
  $F_{Q,\mathrm{proxy}}^{(s)}/(4t^2|\alpha|^2)
  =[\partial_{\Omega}\omega_-^{(s)}]^2$
  as functions of $\Delta_{ad}/\kappa$ at
  $\delta_{ao}^{(0)}/\kappa=-1,0,1$.
  Solid and dashed curves denote the LD and RD branches,
  respectively.
  (b)~Logarithmic directional contrast
  $\log_{10}[F_{Q,\mathrm{proxy}}^{\rm LD}/
  F_{Q,\mathrm{proxy}}^{\rm RD}]$
  in the same parameter plane.
  Gray regions denote parameter points where the directional ratio
  is not evaluated because at least one branch is excluded by the
  validity conditions of the effective model.}
  \label{fig:QFI_nonreciprocal}
\end{figure}

The line cuts and directional ratio in
Fig.~\ref{fig:QFI_nonreciprocal} directly quantify this
nonreciprocal contrast. Along the representative cuts and over most
of the common valid parameter region, the RD response is larger
than the LD, exhibiting the stronger proxy QFI response. The favored direction is determined by the
chosen operating point and is not a universal property of the two drive labels.

\begin{figure}[!htb]
  \centering
  \includegraphics[width=0.72\columnwidth]{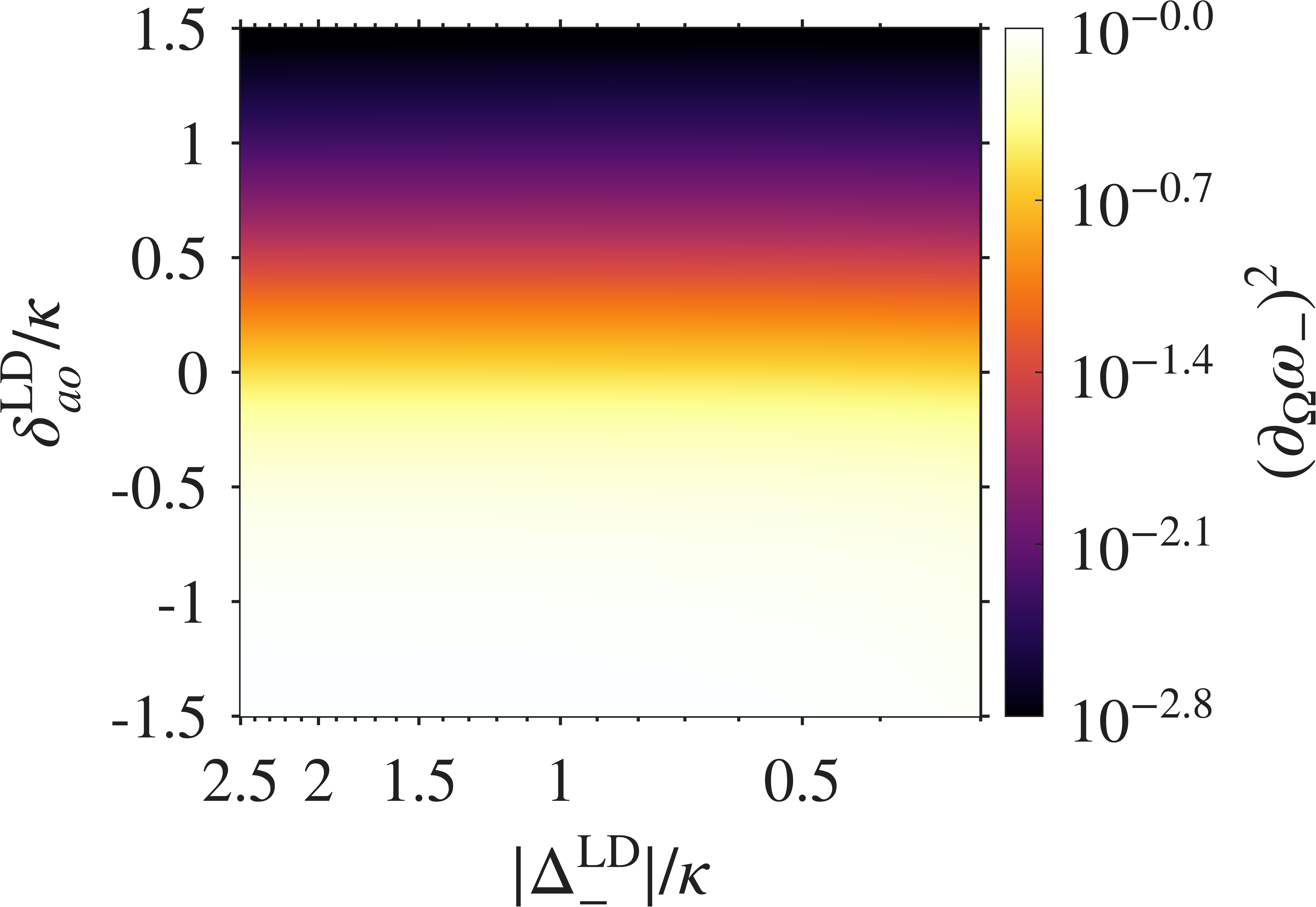}
  \caption{Joint influence of the SW denominator distance and
  polariton hybridization on the lower-polariton proxy QFI within
  the valid effective-model regime.
  The color scale shows
  $F_{Q,\mathrm{proxy}}^{\rm LD}/(4t^2|\alpha|^2)
  =[\partial_{\Omega}\omega_-^{\rm LD}]^2$
  as a function of the LD dressed-state denominator distance
$|\Delta_-^{\rm LD}|/\kappa$ and the renormalized cavity--bosonic-mode detuning
$\delta_{ao}^{\rm LD}/\kappa=(A_b^{\rm LD}-\omega_o)/\kappa$.}
  \label{fig:QFI_denominator_path}
\end{figure}

To separate the role of the microscopic SW denominator from that of the observable polariton composition, Fig.~\ref{fig:QFI_denominator_path}
resolves the LD-branch proxy QFI in terms of
$|\Delta_-^{\rm LD}|$ and $\delta_{ao}^{\rm LD}$. The horizontal
axis characterizes the SW-denominator distance, while the vertical axis characterizes the renormalized cavity--bosonic-mode detuning that controls the lower-polariton composition. The LD branch is used as a representative directional path; the same construction applies to the RD branch. For the plotted convention, the bright lower region corresponds to the predominantly cavity-like side of the lower
polariton, which is more sensitive to the rotation-induced cavity frequency shift, whereas the dark upper region corresponds to the predominantly bosonic-mode-like side with a weaker response. Thus, although the SW denominator introduces the rotation-dependent renormalization of the effective parameters, the observable proxy QFI response is strongly filtered by polariton hybridization. 

\section{\label{sec:experiment}Experimental Implementation and Parameter Estimation}

The experimental platform employs a spinning ring optical
microcavity or an integrated ring resonator~\cite{Maayani2018,Lu2017},
supporting CW and CCW counter-propagating modes
$a_{\rm cw}$ and $a_{\rm ccw}$. A directional coupler selectively
injects LD or RD input fields. Under driving, the cavity field forms
a linearized coupling with a bosonic mode
$o\in\{b,m\}$~\cite{Aspelmeyer2014,LachanceQuirion2019}, as described by
Eq.~\eqref{eq:Heff}. Controllable experimental parameters include
the drive frequency and power, the TLS drive strength $\xi$, the
optical linewidth $\kappa$, the bosonic mode damping rate
$\gamma_o$, and the rotation angular velocity $\Om$.

Although $\DelF(\Om)$ is given analytically by
Eq.~\eqref{eq:fizeau_mag}, experimental calibration is required
for a real device~\cite{Maayani2018,Lu2017}. One may first suppress
the mediated interaction or operate in a far-detuned weak coupling
regime to track the bare cavity resonance. The CW and CCW
transmission spectra are then measured at a series of angular
velocities $\Om_k$, yielding the resonance frequencies
$\hat{\omega}_{a,\rm LD}(\Om_k)$ and
$\hat{\omega}_{a,\rm RD}(\Om_k)$. The calibrated Sagnac-Fizeau shift is
obtained from
\begin{equation}
\hat{\Delta}_F(\Om_k)
=
\frac{1}{2}
\left[
\hat{\omega}_{a,\rm LD}(\Om_k)
-
\hat{\omega}_{a,\rm RD}(\Om_k)
\right],
\end{equation}
and may be fitted in the operating range as
$\DelF(\Om)\simeq k_F\Om$.

The metrological operating point should be chosen where
$|\partial_\Om\omega_-^{(s)}|$ is large while the readout noise
remains controllable and all effective-model validity conditions
remain satisfied. Figures~\ref{fig:QFI_heatmap} and
\ref{fig:QFI_nonreciprocal} identify direction-dependent
high-sensitivity regions in the
$(\Delta_{ad},\delta_{ao}^{(0)})$ parameter plane, while
Fig.~\ref{fig:QFI_denominator_path} illustrates how the microscopic
SW-denominator dependence is converted into an observable
lower-polariton response only after polaritonic hybridization is
taken into account. In general, Eq.~\eqref{eq:dw_result_app} shows
that the measured spectral response contains both the
renormalized frequency channel and the effective-coupling channel;
their relative importance is determined by the selected operating
point.

Two experimentally accessible readout paths can be considered.

\textbf{A. Spectral readout (primary readout port).}---
One may perform spectral measurement or phase-locked tracking of
the lower-polariton branch, using the estimated resonance frequency
$y=\omega_-^{(s)}$ as the signal, or equivalently monitor a
quadrature response associated with the driven lower-polariton mode.
This readout directly corresponds to the lower-polariton proxy QFI
analysis in Sec.~\ref{sec:qfi}; the corresponding exact two-mode
Gaussian QFI is given in Appendix~\ref{app:gaussian_qfi}.
Near an operating point, error propagation gives~\cite{Degen2017}
\begin{equation}
\sigma_\Om\simeq\frac{\sigma_y}{|\partial_\Om y|},
\end{equation}
so that enhancement of $|\partial_\Om\omega_-|$ directly reduces
the estimation uncertainty.

\textbf{B. Bundle-emission readout (auxiliary readout port).}---
When a direction-selective bundle-emission resonance is satisfied,
one may monitor the bundle-event rate
$\hat{\Gamma}_{\rm bun}=N_{\rm bun}/T$ or a zero-delay
two-quantum bundle-correlation signal
$g^{(2)}_{2,ao}(0)$, with $o=b$ or $m$, as an auxiliary
readout~\cite{PRL_NonreciprocalBundle_2024}.
Since the resonance condition depends explicitly on $\DelF(\Om)$,
varying $\Om$ shifts the system toward or away from the
direction-selective resonance window, producing an additional
direction-sensitive signal. This auxiliary channel is not required
for the proxy QFI mechanism above, but may provide a useful
experimental indicator of the rotation-induced nonreciprocity.

Finally, a differential readout may be constructed as
$\hat{y}_\Delta\equiv\hat{y}_{\rm LD}-\hat{y}_{\rm RD}$, with
$y_\Delta(\Om)=y_{\rm LD}(\Om)-y_{\rm RD}(\Om)$,
in analogy with differential Sagnac and dual-interferometer
rotation readout schemes~\cite{Gustavson2000,Moan2020,Tackmann2014}. 
If the LD and RD responses are measured in the same device under interleaved or otherwise well-calibrated conditions, pump-power drift, detection-gain drift, and slow temperature drift can contribute correlated common-mode noise and may be partially suppressed in $y_\Delta$~\cite{Gustavson2000,Tackmann2014}. At the same time, the direction-dependent slopes generated by the opposite Sagnac-Fizeau shifts can enhance $|\partial_\Om y_\Delta|$ near an appropriate operating point~\cite{Post1967,Maayani2018,Lu2017}. The corresponding error-propagation estimate is
\begin{equation}
\sigma_\Om\simeq
\frac{\sigma_{y_\Delta}}{|\partial_\Om y_\Delta|}.
\end{equation}

\section{\label{sec:conclusion}Conclusion and Outlook}
We have proposed and systematically characterized a nonreciprocal quantum rotation-sensing scheme based on virtual-excitation resources in a
hybrid system. The central physical picture is that
the Sagnac-Fizeau shift $\DelF(\Om)$ simultaneously encodes the
rotation parameter and modifies the effective light--matter Hamiltonian through the virtual-transition energy denominators.
This modification renormalizes the hybrid normal-mode structure,
while the counter-rotating terms in the effective interaction
generate virtual excitations and virtual squeezing intrinsically,
without any externally applied squeezing operation. As a result,
the lower-polariton eigenfrequency acquires an amplified rotation
sensitivity, thereby enhancing the proxy QFI defined from the
lower-polariton spectral response.

In contrast to conventional approaches that rely on engineered
squeezing or entanglement, the metrological resource in our scheme is intrinsic to the strongly coupled hybrid system.
The rotation parameter is encoded through the effective detuning modified by the Sagnac effect, while the virtual-excitation-induced
renormalization amplifies its imprint on the lower-polariton
frequency. Nonreciprocity plays an essential role in this picture:
because the LD and RD drives experience opposite Sagnac-Fizeau
shifts, they trace distinct trajectories in the effective
Hamiltonian parameters, including $\Delta_s(\Om)$,
$\tilde{\omega}_a^{(r,s)}(\Om)$, and
$g_{\rm eff}^{(s)}(\Om)$. This asymmetry produces a
direction-dependent proxy QFI contrast, enabling operating-point
optimization toward the more sensitive direction and differential
LD/RD readout for rotation-direction discrimination and
common-mode noise suppression.

On the theoretical side, we derived the closed-form expression
for the lower-polariton frequency derivative $d\omega_-/d\Om$,
explicitly resolving the contributions from the
renormalized-frequency channel and the effective-coupling channel.
On the experimental side, the primary readout is spectral tracking
of the lower-polariton branch, while bundle emission monitoring
serves as an auxiliary direction-selective readout channel. The required
ingredients---a spinning ring microresonator, a strongly driven
two-level system, and a bosonic mode---are all within reach of
current cavity QED, optomechanical, and magnonic platforms.

Looking ahead, the scheme can be extended to magnon--cavity,
phonon--cavity, and multimode network
platforms~\cite{Qin2024,Qin2018,Aspelmeyer2014,LachanceQuirion2019}.
Incorporating realistic backscattering, dissipation, and
non-Markovian noise will be important for determining the
practical sensitivity limits of the proposed scheme. More broadly, our results suggest that the interplay between
virtual-excitation resources and nonreciprocal direction selection
may provide a useful design principle for direction-sensitive
quantum sensing in hybrid quantum platforms.

\appendix

\section{\label{app:geff_sw}Schrieffer-Wolff Second-Order
Elimination and the Explicit Structure of $g_{\rm eff}^{(s)}(\Om)$}

This appendix derives $g_{\rm eff}^{(s)}(\Om)$ starting from the full Hamiltonian in Eq.~\eqref{eq:full_H}~\cite{Bravyi2011}.
Note that $r^{(s)}(\Om)$ here is the single-mode cavity-level Bogoliubov squeezing parameter generated by TLS elimination, distinct from the
lower-polariton squeezing parameter $r_-^{(s)}(\Om)$ in Sec.~\ref{sec:qfi}.

\subsection{\label{app:geff_swA}Dressed-state diagonalization}
Defining
$\omega_q=\sqrt{\Delta_{\sigma d}^2+4\xi^2}$ and the rotation angle $\theta$ satisfying $\tan(2\theta)=2\xi/\Delta_{\sigma d}$, the
dressed states are
\begin{equation}
|+\rangle=\cos\theta|g\rangle+\sin\theta|e\rangle,\quad
|-\rangle=\sin\theta|g\rangle-\cos\theta|e\rangle.
\end{equation}
In the dressed basis, $H_\sigma=(\omega_q/2)\tau_z$, and the
cavity-TLS interaction decomposes as
\begin{align}
H_{a\sigma}=\lambda_{a\sigma}\!\Big[&g_R(a\tau_++a^\dagger\tau_-)
-g_{CR}(a\tau_-+a^\dagger\tau_+)\nonumber\\
&+g_z(a+a^\dagger)\tau_z\Big],
\end{align}
where $g_R=\sin^2\theta$, $g_{CR}=\cos^2\theta$, $g_z=\sin\theta\cos\theta$. And $\tau_+=|+\rangle\langle -|$, $\tau_-=|-\rangle\langle +|$, $\tau_z=|+\rangle\langle +| - |-\rangle\langle -|$.

\subsection{\label{app:geff_swB}Schrieffer-Wolff transformation and effective quadratic term}
For a fixed drive direction $s=+1$ ($-1$) for LD (RD), we
use the direction-dependent detuning
$\Delta_s(\Om)=\Delta_{ad}+s\DelF(\Om)$ and the virtual-transition
denominators $\Delta_\pm(\Om)$ defined in
Eq.~\eqref{eq:delta_pm_main}. After eliminating the
flip terms and projecting onto the $|-\rangle$ branch within the dispersive regime $\left|\Delta_\pm(\Om)\right|
\gg \lambda_{a\sigma}$, the effective cavity mode Hamiltonian is:
\begin{equation}
H_{a,{\rm eff}}^{(s)}(\Om)=
\tilde\omega_a^{(s)}(\Om)\,a^\dagger a
+\frac{\chi^{(s)}(\Om)}{2}(a^2+a^{\dagger2}),
\end{equation}
where
\begin{align}
\tilde\omega_a^{(s)}(\Om)
&=\Delta_s(\Om)+\lambda_{a\sigma}^2\!\left[
\frac{g_R^2}{\Delta_-(\Om)}-\frac{g_{CR}^2}{\Delta_+(\Om)}
\right],
\label{eq:omega_a_tilde_app}\\
\chi^{(s)}(\Om)
&=-\lambda_{a\sigma}^2\,g_Rg_{CR}\!\left[
\frac{1}{\Delta_-(\Om)}-\frac{1}{\Delta_+(\Om)}
\right].
\label{eq:chi_def}
\end{align}
Here, the term
$\frac{\chi^{(s)}(\Om)}{2}(a^2+a^{\dagger2})$
is a TLS-mediated single-mode counter-rotating term generated by
the SW elimination. It produces a virtual single-mode squeezing
structure without any externally applied squeezing drive, and thus
contributes to the virtual-excitation resources of the effective
cavity mode~\cite{DeLiberato2017,Ciuti2005}.

\subsection{\label{app:geff_swC}From virtual squeezing to $g_{\rm eff}^{(s)}(\Om)$}
The single-mode Bogoliubov transformation~\cite{Weedbrook2012}
\begin{equation}
a=\cosh{[r^{(s)}]}\tilde{a}+\sinh{[r^{(s)}]}\tilde{a}^\dagger
\label{eq:bogoliubov_transform}
\end{equation}
eliminates the $\tilde{a}^2+\tilde{a}^{\dagger2}$ terms when
\begin{equation}
\tanh{[2r^{(s)}(\Om)]}=
-\frac{\chi^{(s)}(\Om)}{\tilde\omega_a^{(s)}(\Om)}.
\label{eq:tanh_condition}
\end{equation}
The diagonalized single cavity Hamiltonian is
$\tilde\omega_a^{(r,s)}\,\tilde{a}^\dagger\tilde{a}$ with
\begin{equation}
\tilde\omega_a^{(r,s)}=
\frac{\tilde\omega_a^{(s)}}{\cosh{[2r^{(s)}]}},
\end{equation}
and the quadrature rescaling $a+a^\dagger$
gives
\begin{equation}
a+a^\dagger=e^{r^{(s)}(\Om)}\,(\tilde{a}+\tilde{a}^\dagger).
\label{eq:quadrature_rescaling}
\end{equation}
Substituting into $\lambda_{ao}(a+a^\dagger)(o+o^\dagger)$ yields
\begin{equation}
g_{\rm eff}^{(s)}(\Om)\,(\tilde{a}+\tilde{a}^\dagger)(o+o^\dagger),
\qquad
g_{\rm eff}^{(s)}(\Om)=\lambda_{ao}\,e^{r^{(s)}(\Om)}.
\label{eq:geff_sw_final}
\end{equation}
Since $r^{(s)}(\Om)$ depends on $\Delta_\pm(\Om)$ through
$\chi^{(s)}/\tilde\omega_a^{(s)}$, the LD and RD values of
$g_{\rm eff}^{(s)}$ are generally different. This constitutes one
part of the direction-dependent effective-parameter renormalization;
the observable sensitivity is determined after combining this
coupling renormalization with the renormalized cavity frequency and
the lower-polariton hybridization.

\section{\label{app:diag_and_derivative}Explicit Derivation of Polariton Diagonalization and the Sensitivity Derivative}

This appendix gives the standard diagonalization procedure for the two-mode Hamiltonian entering the polariton analysis and derives the closed-form lower-polariton sensitivity derivative $d\omega_-/d\Om$ used in the main text. After the SW elimination
(Appendix~\ref{app:geff_sw}) and the subsequent single-mode
Bogoliubov transformation, the cavity mode $a$ is replaced by
$\tilde{a}$ and the cavity frequency is further renormalized to
$\tilde\omega_a^{(r,s)}=\tilde\omega_a^{(s)}/\cosh(2r^{(s)})$.
The two-mode Hamiltonian entering the polariton diagonalization is
therefore
\begin{equation}
H_{\rm eff}^{(s)}(\Om)=\tilde{\omega}_a^{(r,s)}(\Om)\,\tilde{a}^\dagger\tilde{a}
+\omega_o\,o^\dagger o
+g_{\rm eff}^{(s)}(\Om)(\tilde{a}+\tilde{a}^\dagger)(o+o^\dagger).
\end{equation}
Throughout this appendix, the interaction is written as
$(\tilde{a}+\tilde{a}^\dagger)(o+o^\dagger)$, which is why the
discriminant contains the coefficient~$16$.

\subsection{\label{app:diag_and_derivativeA}Polariton eigenfrequencies}
Introducing the shorthand
\begin{equation}
A(\Om)\equiv\tilde{\omega}_a^{(r,s)}(\Om),\qquad
B\equiv\omega_o,\qquad
g(\Om)\equiv g_{\rm eff}^{(s)}(\Om),
\end{equation}
the standard Hopfield/Bogoliubov symplectic
diagonalization~\cite{Hopfield1958} yields two polariton modes
$\hat{e}$ and $\hat{f}$ with positive eigenfrequencies
\begin{align}
\omega_\pm(\Om)=\Bigg[\frac{1}{2}\Bigg(
&A^2+B^2\nonumber\\
&\pm\sqrt{(A^2-B^2)^2+16g^2 AB}
\Bigg)\Bigg]^{1/2},
\label{eq:omega_pm_app}
\end{align}
with auxiliary quantities
\begin{align}
D(\Om)&=\sqrt{(A^2-B^2)^2+16g^2 AB},\\
Y(\Om)&=\frac{1}{2}(A^2+B^2-D),
\end{align}
so that $\omega_-(\Om)=\sqrt{Y(\Om)}$.

\subsection{\label{app:diag_and_derivativeB}Derivative of the lower branch with respect to $\Om$}
By the chain rule,
\begin{equation}
\frac{d\omega_-}{d\Om}
=\frac{1}{2\omega_-}\frac{dY}{d\Om},
\qquad
\frac{dY}{d\Om}=AA'-\frac{1}{2}D',
\end{equation}
where $A'\equiv dA/d\Om$ and $g'\equiv dg/d\Om$.
For $D=\sqrt{Q}$ with $Q=(A^2-B^2)^2+16g^2AB$,
one has $D'=Q'/(2D)$, where
\begin{equation}
Q'=4A(A^2-B^2)A'+32gg'AB+16g^2BA'.
\end{equation}
Rearranging gives
\begin{equation}
\frac{dY}{d\Om}
=
AA'
-\frac{A'}{D}\left[A(A^2-B^2)+4g^2B\right]
-\frac{8gg'AB}{D},
\end{equation}
and substituting into $d\omega_-/d\Om=(2\omega_-)^{-1}dY/d\Om$ yields
\begin{equation}
\frac{d\omega_-}{d\Om}
=
\frac{1}{2\omega_-}
\left[
AA'
-\frac{A'}{D}\left(A(A^2-B^2)+4g^2B\right)
-\frac{8gg'AB}{D}
\right].
\label{eq:dw_result_app}
\end{equation}
This is the closed-form lower-polariton sensitivity derivative $d\omega_-/d\Om$ used in the main text, with
$A\equiv\tilde\omega_a^{(r,s)}(\Om)$ and
$g\equiv g_{\rm eff}^{(s)}(\Om)$ carrying an implicit direction index $s=\pm1$ in the direction-dependent implementation.
The derivative $A'\equiv d\tilde\omega_a^{(r,s)}/d\Om$ depends on
$\tilde\omega_a^{(s)}$, $\chi^{(s)}$, and $r^{(s)}$ through the
Bogoliubov relation in Eq.~\eqref{eq:tanh_condition}. It is evaluated numerically, instead of reducing to
a simple closed form.

\subsection{\label{app:diag_and_derivativeC}Nonlinear $\Om$ dependence: effective coupling derivative}
The effective coupling $g_{\rm eff}^{(s)}(\Om)$ satisfies
\begin{equation}
g_{\rm eff}^{(s)}(\Om)=\lambda_{ao}
\left(\frac{\tilde{\omega}_a^{(s)}(\Om)-\chi^{(s)}(\Om)}
{\tilde{\omega}_a^{(s)}(\Om)+\chi^{(s)}(\Om)}\right)^{1/4},
\label{eq:geff_explicit}
\end{equation}
with the derivative
\begin{equation}
g_{\rm eff}^{(s)\prime}(\Om)=\frac{g_{\rm eff}^{(s)}(\Om)}{4}
\left[\frac{\tilde{\omega}_a^{(s)\prime}-\chi^{(s)\prime}}
{\tilde{\omega}_a^{(s)}-\chi^{(s)}}
-\frac{\tilde{\omega}_a^{(s)\prime}+\chi^{(s)\prime}}
{\tilde{\omega}_a^{(s)}+\chi^{(s)}}\right].
\label{eq:geff_prime}
\end{equation}
Here, $\tilde{\omega}_a^{(s)}(\Om)$ and $\chi^{(s)}(\Om)$ are given by Eqs.~\eqref{eq:omega_a_tilde_app}--\eqref{eq:chi_def} in Appendix~\ref{app:geff_sw}, and the $\Om$ argument is suppressed for brevity. Equation~\eqref{eq:geff_prime} shows explicitly that $g_{\rm eff}^{(s)\prime}(\Om)$ inherits nonlinear $\Om$ dependence from the Sagnac-shifted energy denominators $\Delta_\pm(\Om)$.
This effective-coupling derivative provides one contribution to the direction-asymmetric lower-polariton response, together with the renormalized frequency derivative
$\partial_\Om\tilde{\omega}_a^{(r,s)}$ in
Eq.~\eqref{eq:dw_result_app}.

\section{\label{app:critical_validation}Numerical Validation of the Near-Critical Approximation}

To verify the range of validity of the near-critical approximation
used in Eq.~\eqref{eq:FQ_exp_qfi}, we compare it with the exact
lower-polariton derivative used in the proxy QFI calculation,
obtained from Eq.~\eqref{eq:dw_result_app}. Specifically, we define
\begin{equation}
F_{\rm exact}^{(s)}
=
\left[\partial_\Omega\omega_-^{(s)}\right]^2,
\end{equation}
where $\partial_\Omega\omega_-^{(s)}$ is evaluated from the full
closed-form derivative, and
\begin{equation}
F_{\rm app}^{(s)}
=
\frac{\left[\mathcal W^{(s)}\right]^2
\left[\partial_\Omega q_s\right]^2}
{4\left(1-q_s\right)}
\end{equation}
from the leading near-critical expression. Since this approximation
retains only the leading singular term as $q_s\to1^-$, it is not
expected to reproduce the full valid parameter region away from the
critical threshold.

\begin{figure}[t]
  \centering
  \includegraphics[width=0.9\columnwidth]{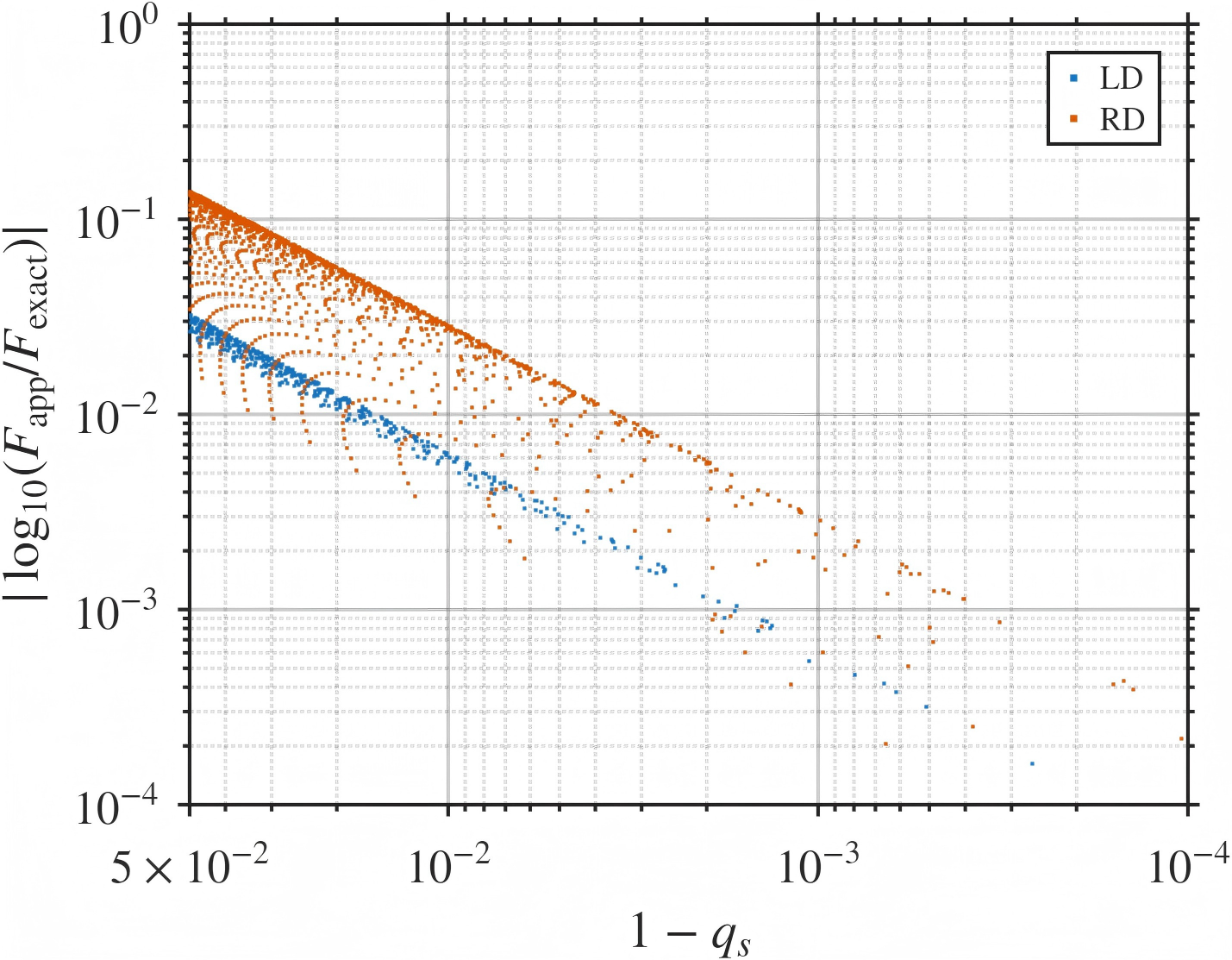}
  \caption{
  Validation of the near-critical asymptotic expression.
  We compare the approximate result $F_{\rm app}$ with
  $F_{\rm exact}=(\partial_\Omega\omega_-^{(s)})^2$, where
  $\partial_\Omega\omega_-^{(s)}$ is evaluated from the exact
  lower-polariton derivative used in the proxy QFI calculation. The logarithmic deviation $|\log_{10}(F_{\rm app}/F_{\rm exact})|$ is plotted versus $1-q_s$ for the LD and RD branches. Only points with
  $10^{-4}\le 1-q_s\le 5\times10^{-2}$ are shown. The deviation
  decreases systematically as $q_s\to1^-$, confirming the validity
  of the leading asymptotic expression in the near-critical regime.
  }
  \label{fig:critical_validation}
\end{figure}

As shown in Fig.~\ref{fig:critical_validation}, the logarithmic
deviation decreases as the critical distance $1-q_s$ becomes
smaller. This confirms that Eq.~\eqref{eq:FQ_exp_qfi} is a genuine
near-critical asymptotic result rather than a global approximation
over the full stable parameter space.
\section{\label{app:gaussian_qfi}Gaussian-State QFI, Proxy Approximation, and Single-Polariton Limit}

The effective Hamiltonian in Eq.~\eqref{eq:Heff} is quadratic, and an initial Gaussian state remains Gaussian under the symplectic evolution $S(\Om,t)=\exp[JG(\Om)t]$ ($J$ is the symplectic form matrix, $G(\Om)$ is the quadratic-form matrix)~\cite{Weedbrook2012}. The first moment $\bm{d}$ and covariance matrix $V$ evolve as
\begin{equation}
\bm{d}(\Om,t)=S(\Om,t)\bm{d}_0,\qquad V(\Om,t)=S\,V_0\,S^T.
\end{equation}
The \emph{exact QFI} of a Gaussian state is given by the general covariance matrix formula~\cite{Safranek2018,Bakmou2020,Liu2019}:
\begin{align}
F_Q(\Om)=&\;\frac{1}{2}{\rm Tr}\!\left[(\partial_\Om V)\!\left(V+\frac{i}{2}J\right)^{\!-1}(\partial_\Om V)\!\left(V-\frac{i}{2}J\right)^{\!-1}\right]\nonumber\\
&+(\partial_\Om\bm{d})^T V^{-1}(\partial_\Om\bm{d}).
\label{eq:QFI_Gaussian}
\end{align}

\subsection{\label{app:gaussian_qfiA}Proxy approximation}
In the lower-polariton-dominated coherent driving regime considered in the main text, the first-moment response $\bm{d}(\Om,t)$ is predominantly determined by the lower-polariton branch, so that $(\partial_\Om\bm{d})^T V^{-1}(\partial_\Om\bm{d})\approx 4t^2|\alpha|^2(\partial_\Om\omega_-)^2$. Furthermore, when $|\alpha|^2\gg1$, the covariance matrix term $F_Q^{(V)}$ is $O(1)$ while the first-moment term $F_Q^{(d)}$ is $O(|\alpha|^2)$, so the former is negligible and Eq.~\eqref{eq:QFI_Gaussian} reduces
to the proxy QFI expression in Eq.~\eqref{eq:FQ_proxy}. For arbitrary parameter ranges, Eq.~\eqref{eq:QFI_Gaussian} provides the exact Gaussian QFI: constructing $G(\Om)\to S(\Om,t)\to V(\Om,t)\to F_Q(\Om)$ and taking $\DelF=\pm|\DelF|$ respectively yields $F_Q^{\rm LD}(\Om)$ and $F_Q^{\rm RD}(\Om)$ for directional comparison.

\subsection{\label{app:gaussian_qfiB}Single-polariton limit}
If one further adopts the reduced single-polariton description $H_-(\Om)=\omega_-(\Om)\hat{e}^\dagger\hat{e}$ with a coherent initial state and unitary evolution, the covariance matrix becomes $\Om$-independent and the first term in Eq.~\eqref{eq:QFI_Gaussian} vanishes exactly, making Eq.~\eqref{eq:FQ_proxy} an exact result rather than an approximation in that more restricted framework~\cite{PRL_USC_Metrology_2025}.

\begin{acknowledgments}
This work is supported by the National Natural Science Foundation of China under Grants No. W2411002 and No. 12375018.
\end{acknowledgments}

\bibliographystyle{apsrev4-2}
\nocite{apsrev42Control}
\bibliography{refs}

\end{document}